# A coarse-grid projection method for accelerating incompressible MHD flow simulations

Ali Kashefi[1, 2]

kashefi@stanford.edu

1: Department of Mechanical Engineering, Stanford University, Stanford, CA 94305, USA

2: Department of Civil and Environmental Engineering, Stanford University, Stanford, CA 94305, USA

## Abstract

Coarse grid projection (CGP) is a multiresolution technique for accelerating numerical calculations associated with a set of nonlinear evolutionary equations along with stiff Poisson's equations. In this article we use CGP for the first time to speed up incompressible magnetohydrodynamics (MHD) flow simulations. Accordingly, we solve the nonlinear advection-diffusion equation on a fine mesh, while we execute the electric potential Poisson equation on the corresponding coarsened mesh. Mapping operators connect two grids together. A pressure correction scheme is used to enforce the incompressibility constrain. The study of incompressible flow past a circular cylinder in the presence of Lorentz force is selected as a benchmark problem with a fixed Reynolds number but various Stuart numbers. We consider two different situations. First, we only apply CGP to the electric potential Poisson equation. Second, we apply CGP to the pressure Poisson equation as well. The maximum speedup factors achieved here are approximately 3 and 23 respectively for the first and second situations. For the both situations we examine the accuracy of velocity and vorticity fields as well as the lift and drag coefficients. In general, the results obtained by CGP are in an excellent to reasonable range of accuracy and are significantly consistently more accurate than when we use coarse grids for the discretization of both the advection-diffusion and electric potential Poisson equations.

# 1 Introduction

The study of external and internal [12] magnetohydrodynamics (MHD) flows has been the subject of interest in scientific areas (see e.g., Mutschke, et al. [19]; Klein and Pothérat [15]; Farooq, et al. [4]) as well as industrial fields (see e.g., Josserand, et al. [10]; Irfan and Farooq [8]; Shoaib, et al. [26]). Numerous researchers have taken advantages of computational tools for the MHD flow simulations in desired conditions (Verron and Sommeria [27]; Mück, et al. [18]; Yoon, et al. [28]; Lee, et al. [16]; Dousset and Pothérat [3]; Shah, et al. [25]; Peng, et al. [20]). From a mathematical perspective, one has to deal with a Poisson equation for an electric potential field in order to obtain a numerical solution to the solenoidal current field governed by the Ohm law (see e.g., Sect. 2 of Lee, et al. [16]). The challenge ahead is the fact that the elliptic Poisson equation is the most time consuming component of the numerical procedure and demands high computational expenses. An approach to substantially lessen the central processing unit (CPU) times associated with the linear electric potential Poisson equation is using multi-resolution techniques. To this purpose, we examine one of the recently introduced techniques, called Coarse Grid Projection (CGP) methodology (Lentine, et al. [17]; San and Staples [23]; San and Staples [24]; Jin, et al. [9]; Kashefi and Staples [14]; Kashefi [13]; Kashefi [11]), for the MHD flow computations for the first time in this research paper.

Historically, the CGP algorithm was presented first by Lentine, et al. [17], when they applied this multi-resolution method to inviscid flow simulations for video games. San and Staples [23] implemented CGP in numerical computations of transient incompressible viscous flows for high Reynolds numbers. Furthermore, they used CGP for quasigeostrophic ocean simulations (San and Staples [24]). Jin, et al. [9] calibrated the fast fluid dynamics (FFD) models using CGP. Kashefi and Staples [14] proposed a specific version of CGP for unstructured triangular finite element grids. Kashefi [13] discussed CGP as a partial mesh refinement tool for the incompressible flow simulations. Kashefi [11] constructed a CGP framework for incremental pressure correction schemes. Lastly, the effect of CGP on the energy equation was studied by Kashefi [12].

Generally, calculations relevant to the velocity field are performed on a grid with fine spatial resolutions, mainly due to the existence of the nonlinear convection term in the Navier-Stokes equations. On the other hand, the velocity field appears in the right hand side of the electric potential Poisson equation in the MHD flows. By applying the CGP approach to this set of equations, instead of solving the Poisson equation on the fine mesh, the velocity field is restricted onto a corresponding coarsened grid and the potential Poisson equation is solved on the coarsened grid. Since the degree of freedom of Poisson's equation decreases, a considerable amount of CPU times is saved. The resulting electric potential is prolonged onto the fine grid and the Lorentz force is set as the source term of the Navier-Stokes equations. Note that Kashefi and Staples [14] demonstrated that the CGP approach preserves the accuracy of the pressure gradient field, instead of the pressure itself. On the other hand, the pressure gradient appears as an explicit variable in the Navier-Stokes equations. Hence, the velocity field fidelity is preserved in CGP as well. Making an analogy, the electric potential is also an implicit variable in the Lorentz force, which means that we deal with the gradient of the electric potential field, instead of the electric potential itself. Thus, it is expected that CGP preserves the accuracy of the Lorentz force, and consequently, the accuracy of the velocity field. It is worthwhile to note that we use a non-incremental pressure projection scheme to execute the incompressibility constraint of the velocity field. As a result, at each time step, a set of three decoupled elliptic equations has to be numerically solved: a linear electric potential Poisson equation, a nonlinear advection-diffusion equation, and a linear pressure Poisson equation. We investigate the integrity of the unsteady incompressible MHD flows when CGP is only applied to the potential Poisson equation or simultaneously to both the potential and pressure Poisson equations.

The study of the flow around bluff bodies has been long undertaken for scientific purposes as well as engineering applications (Zhao and Lu [29]; Ganta, et al. [5]). As regards the MHD flows, numerous experimental and computational observations for the fluid flow past an obstacle have been reported in the literature (Mutschke, et al. [19]; Yoon, et al. [28]; Dousset and Pothérat [3]; Ghosh, et al. [7]). Due to the importance of this fundamental fluid mechanics problem, we peak the two-dimensional incompressible MHD flows around a circular cylinder as the benchmark test case for the present numerical study. We perform the numerical simulations for a fixed Reynolds number and different values of the Stuart number. For each case, we evaluate the CPU time, lift and drag evolutions, and von Karman vortex street structure for the CGP versus non-CGP computations. We use the most generalized version of the CGP framework introduced by Kashefi and Staples [14]. In this way, we examine the CGP performance for the electric potential Poisson equation in an unstructured triangular finite element mesh.

This article is structured as follows. We present the governing equations of the flow field in the presence of a magnetic field in Sect. 2.1. Coarse grid projection methodology is provided in Sect. 2.2. Computational aspects of the present study are discussed in Sect 2.3. Resulting simulations with graphical and numerical explanations are illustrated in Sect. 3. Finally, conclusions of the work are given in Sect. 4.

## 2 Problem formulation

### 2.1 Governing equations

We consider an incompressible isothermal flow affected by the Lorentz force. Hence, the equations of the conservation of charge, momentum, and mass govern the fluid flow:

$$\nabla \cdot \boldsymbol{j} = 0, \tag{1}$$

$$\rho \left[ \frac{\partial \boldsymbol{u}}{\partial t} + (\boldsymbol{u} \cdot \nabla)\boldsymbol{u} \right] - \mu \Delta \boldsymbol{u} + \nabla p = \boldsymbol{j} \times \boldsymbol{B}, \tag{2}$$

$$\nabla \cdot \boldsymbol{u} = 0, \tag{3}$$

where $\boldsymbol{u}$, $\boldsymbol{j}$, and $\boldsymbol{B}$ denote the vector of the velocity, the current density, and the static magnetic field, respectively. $p$ stands for the pressure. $\rho$ represents the fluid flow density and $\mu$ indicates the dynamic viscosity.

Ohm's law [1] says:

$$\boldsymbol{j} = \sigma(-\nabla \phi + \boldsymbol{u} \times \boldsymbol{B}), \tag{4}$$

where $\phi$ is the electric potential and $\sigma$ is the electric conductivity.

Substituting Eq. (4) in Eq. (1), an electric potential Poisson equation is derived:

$$\Delta \phi = \nabla \cdot (\boldsymbol{u} \times \boldsymbol{B}). \tag{5}$$

From a time integration point of view, we explicitly deal with the current density and the Lorentz force, while a semi-implicit first order scheme is used for the velocity and pressure fields. In the next stage, a non-incremental pressure correction scheme (Chorin [2]) is applied to the Navier-Stokes and mass continuity equations. That procedure yields to:

$$\Delta \phi^n = \nabla \cdot (\boldsymbol{u}^n \times \boldsymbol{B}^n), \tag{6}$$

$$\boldsymbol{j}^{n} = \sigma\left(-\nabla \phi^{n} + \boldsymbol{u}^{n} \times \boldsymbol{B}^{n}\right), \tag{7}$$

$$\rho\left[\frac{\tilde{\boldsymbol{u}}^{n+1} - \boldsymbol{u}^{n}}{\delta t} + \left(\boldsymbol{u}^{n} \cdot \nabla\right)\tilde{\boldsymbol{u}}^{n+1}\right] - \mu\Delta\tilde{\boldsymbol{u}}^{n+1} = \boldsymbol{j}^{n} \times \boldsymbol{B}^{n}, \tag{8}$$

$$\Delta p^{n+1} = \frac{\rho}{\delta t}\nabla \cdot \tilde{\boldsymbol{u}}^{n+1}, \tag{9}$$

$$\boldsymbol{u}^{n+1} = \tilde{\boldsymbol{u}}^{n+1} - \frac{\delta t}{\rho}\nabla p^{n+1}, \tag{10}$$

where $\delta t$ indicates the time step and $\tilde{\boldsymbol{u}}^{n+1}$ is the intermediate velocity vector at time $t^{n+1}$.

## 2.2 Coarse grid projection methodology

The CGP scheme potentially is able to accelerate simulations associated with a set of decoupled equations if the following two conditions are satisfied. First, at least one linear equation exists in this set. Second, the unknown variable of the equation must appear implicitly (e.g., its gradient) in other equations of the set. If the conditions are satisfied, then CGP dramatically lowers the computational cost by executing the equation on a coarsened mesh while preserves the accuracy of variables presented explicitly in other equations. For the MHD flows, the electric potential Poisson equation is the linear elliptic equation and the gradient of the electric potential appears in the momentum equation, not the electric potential itself. Hence, these two conditions are satisfied. Note that the Poisson equation is the most time consuming part of the simulation. Figure 1 gives a schematic illustration of the CGP mechanism. As shown in Fig. 1, at each time step $t^n$, we map the velocity field of a fine grid $\boldsymbol{u}_f^n$ onto a corresponding coarsened grid and set $\boldsymbol{u}_c^n$. The restricted velocity field is used to solve Poisson's equation for the electric potential field on the coarsened grid $\phi_c^n$. Then, we remap the resulting potential data $\phi_c^n$ onto the fine grid and set $\phi_f^n$. The gradient of $\phi_f^n$ plays a role in the source term of the momentum equation.

In principle, the CGP scheme described can be potentially applied to three-dimensional MHD flows. In this article, however, we restrict our study to the two dimensional flows in the $x-y$ plane with a constant magnetic field parallel to the $z-$axis such that:

$$\boldsymbol{B} = \mathrm{B}_o \hat{k}, \tag{11}$$

where $B_o$ is the constant. This condition has been applied to the two-dimensional confined jet flows by Lee, et al. [16].

Eqs. (12)-(17) depict the CGP algorithm for the simulation of the two dimensional incompressible MHD flow at each time step.

1. Restrict $u_f^n$ and $v_f^n$ onto the coarsened grid and obtain $u_c^n$ and $v_c^n$.

2. Calculate $\phi_c^n$ on the coarsened mesh by solving

$$\frac{\partial^2 \phi_c^n}{\partial x^2} + \frac{\partial^2 \phi_c^n}{\partial y^2} = B_o \left( \frac{\partial v_c^n}{\partial x} - \frac{\partial u_c^n}{\partial y} \right). \tag{12}$$

3. Prolong $\phi_c^n$ onto the fine grid and obtain $\phi_f^n$.

4. Calculate $\tilde{u}_f^{n+1}$ and $\tilde{v}_f^{n+1}$ on the fine mesh by solving

$$\rho \left( \frac{\tilde{u}_f^{n+1} - \tilde{u}_f^n}{\delta t} + u_f^n \frac{\partial \tilde{u}_f^{n+1}}{\partial x} + v_f^n \frac{\partial \tilde{u}_f^{n+1}}{\partial y} \right) - \mu \left( \frac{\partial^2 \tilde{u}_f^{n+1}}{\partial x^2} + \frac{\partial^2 \tilde{u}_f^{n+1}}{\partial y^2} \right) = \sigma B_o \left( -\frac{\partial \phi_f^n}{\partial y} - B_o u_f^n \right), \tag{13}$$

$$\rho \left( \frac{\tilde{v}_f^{n+1} - \tilde{v}_f^n}{\delta t} + u_f^n \frac{\partial \tilde{v}_f^{n+1}}{\partial x} + v_f^n \frac{\partial \tilde{v}_f^{n+1}}{\partial y} \right) - \mu \left( \frac{\partial^2 \tilde{v}_f^{n+1}}{\partial x^2} + \frac{\partial^2 \tilde{v}_f^{n+1}}{\partial y^2} \right) = \sigma B_o \left( \frac{\partial \phi_f^n}{\partial x} - B_o v_f^n \right). \tag{14}$$

5. Restrict $\tilde{u}_f^{n+1}$ and $\tilde{v}_f^{n+1}$ onto the coarsened grid and obtain $\tilde{u}_c^{n+1}$ and $\tilde{v}_c^{n+1}$.

6. Calculate $p_c^{n+1}$ on the coarse mesh by solving

$$\frac{\partial^2 p_c^{n+1}}{\partial x^2} + \frac{\partial^2 p_c^{n+1}}{\partial y^2} = \frac{\rho}{\delta t} \left( \frac{\partial \tilde{u}_c^{n+1}}{\partial x} + \frac{\partial \tilde{v}_c^{n+1}}{\partial y} \right). \tag{15}$$

7. Prolong $p_c^{n+1}$ onto the fine grid and obtain $p_f^{n+1}$.

8. Calculate $u_f^{n+1}$ and $v_f^{n+1}$ on the fine mesh via

$$u_f^{n+1} = \tilde{u}_f^{n+1} - \frac{\delta t}{\rho} \frac{\partial p_f^{n+1}}{\partial x}, \tag{16}$$

$$v_f^{n+1} = \tilde{v}_f^{n+1} - \frac{\delta t}{\rho}\frac{\partial p_f^{n+1}}{\partial y}. \tag{17}$$

Subscripts $f$ and $c$ indicate the value of the relevant variable on a fine gird and on a corresponding coarsened grid, respectively. $u$ and $v$ are, respectively, the velocity fields in $x$ and $y$ directions, while $\tilde{u}$ and $\tilde{v}$ are, respectively, the intermediate velocity fields in $x$ and $y$ directions. Note that one may omit steps (5)-(7) in the preceding algorithm, which means that CGP is only used for the electric potential Poisson equation.

## 2.3 Computational consideration

The CGP method and its relevant mapping functions depend significantly on the spatial discretization scheme. Here, we use the CGP framework as well as the restriction and prolongation operators introduced by Kashefi and Staples [14] for unstructured finite-element grids. This CGP configuration has been successfully utilized in the literature (Kashefi [11]; Kashefi [13]). Thus, we discretize the system of Eqs. (12)-(17) using the Galerkin finite element method (Reddy [21]) with linear $\mathbf{P}_1$ shape functions. According to the numerical studies investigated in San and Staples [23], Kashefi and Staples [14], and Kashefi [11] CGP was successful for up to three levels of the pressure Poisson grid coarsening. Following the experience, we practically consider four sequences of nested hierarchical grids, which in each fine mesh with $M$ element numbers is generated by uniformly subdividing elements of a coarse grid with $P$ elements. In this way, $M$ and $N$ are correlated together with

$$P = 4^{-k} M, \tag{18}$$

where $k$ is the coarsening level.

We illustrate the restriction and prolongation functions addressed in the steps (1), (3), (5), and (7) of the CGP algorithm. To save space, we describe the mapping technique in a general format. One may apply this scheme to the restriction of the velocity or intermediate velocity fields and to the prolongation of the electric potential or the pressure fields. Let us take a fine finite element space $V_j$ with the data set $\{q_f\}$ resolved on that. The plan is to restrict the data into a coarse finite element space $V_{j-k}$ to obtain the data set $\{q_c\}$. Note that $k$ is a positive integer number indicating the coarsening level as mentioned earlier. Moreover, the dimension of each date set is equal to the total number of degrees of freedom of the corresponding space. Obviously, the dimension of $\{q_f\}$ is greater than $\{q_c\}$. Now consider two random nodes $\alpha$ and $\beta$ which belong, respectively, to spaces $V_j$ and $V_{j-k}$. We enforce

$$q_c(\beta) = q_f(\alpha) \text{ if } \mathbf{x}(\beta) = \mathbf{x}(\alpha), \tag{19}$$

where $\mathbf{x}(\beta)$ is the position vector of node $\beta$ on the space $V_{j-k}$. $\mathbf{x}(\alpha)$ is the position vector of node $\beta$ on the space $V_j$. similarly defined. Notice that all finite element spaces are located in one coordinate system. In other words, we simply use an injection procedure to perform the restriction. Taking the advantage of this interpolation, one may directly restrict the data from a desired space onto another non-nested space, which reduces the associated computational cost. Now we explain the prolongation platform. The goal is to prolong the data set $\{q_c\}$ of the coarse space $V_{j-k}$ onto the fine space $V_j$ to set $\{q_f\}$. In finite element applications, a possible strategy for the construction of the prolongation operator is the implementation of the finite element shape function. In the current study, we use linear shape functions. Thus, we prolong the data linearly. For any node $\alpha$ on $V_{j-k+1}$, there exists a linear triangular element $E$ with three nodes $\beta_j$ ($1 \le j \le 3$) in the space of $V_{j-k}$ such that $\mathbf{x}_\alpha \in V_E \subset V_{j-k}$. Thus $q_f(\alpha)$ can be computed as a linear combination of the nodal values of $V_E$ such that

$$q_f(\alpha) = \sum_{j=1}^{3} q_c(\beta_j) \psi_j(\mathbf{x}_\alpha), \tag{20}$$

where $\psi_j$ is a linear shape function of element $E$. In contrast with the restriction function, we can only prolong the data from a desired space $V_{j-k}$ to the nested space $V_{j-k+1}$. For instance, one has to run the prolongation function three times for a simulation with three levels of coarsening $(k=3)$. That is why the restriction operator is more costly in comparison with the prolongation operator. One may see Sect. 2.3 of Kashefi and Staples [14] for further elaboration. The efficiency of CGP strongly depends on the design of the restriction and prolongation operators. However, we demonstrate that the CGP methodology is quite proficient even using these basic data interpolation/extrapolation.

An object oriented C++ code is developed. The GMRES(m) technique with ILU(0) preconditioner [22] is used to numerically solve Eqs. (12)-(15). Gmsh [6] is used as the finite element mesh generator. All simulations are performed on a single Intel(R) Xeon(R) processor with 2.66 GHz clock rate and 64 Gigabytes of RAM.

## 3 Results and discussion

A rectangular box $[0,L]\times[0,H]$ is considered as the computational domain. A circle with a diameter of $d$ and center of $(x_c, y_c)$ represents a rigid circular cylinder with no-slip conditions. We generate the meshes with 108352 nodes and 215680 elements, 27216 nodes and 53920 elements, 6868 nodes and 13480 elements, and 1749 nodes and 3370 elements, respectively, for $k=0$, $k=1$, $k=2$, and $k=3$. We show those grids for $k=2$ and $k=3$ in Fig. 2.

The inflow boundary condition is modeled by a free stream velocity

$$\boldsymbol{u} = u_\infty \hat{\boldsymbol{i}}, \tag{21}$$

and the outflow by the natural Neumann condition

$$\mu \nabla \boldsymbol{u} \cdot \boldsymbol{n} = 0. \tag{22}$$

The bottom and top of the domain are perfectly slipped. For the magnetic field we follow the boundary conditions described in Refs. [18, 3, 19]. Accordingly, far from the cylinder, the electric field vanishes.

The Reynolds number is determined as

$$\mathrm{Re} = \frac{\rho d u_\infty}{\mu}. \tag{23}$$

In this study, we set this dimensionless number to $\mathrm{Re}=100$. In the International Unit System, we set the density $(\rho)$, cylinder diameter $(d)$, and free stream velocity $(u_\infty)$ to 1.00; and the viscosity $(\mu)$ to 0.01. The length $(L)$ and the height $(H)$ of the rectangular box are, respectively, equal to 38 m and 32 m. For the center of the circle we set $x_c = 8m$ and $y_c = 16m$.

The Stuart number or the magnetic interaction parameter is expressed as

$$N = \frac{\sigma B_o^2 d}{\rho u_\infty}. \tag{24}$$

The constant magnetic field ($B_o$) is set to 1.00 in the International Unit System. We vary the electric conductivity ($\sigma$) of the fluid to set the Stuart number. The numerical computations are performed until time $t = 150\,\text{s}$ with the fixed time increment of $\delta t = 0.05\,\text{s}$.

We demonstrate the grid resolution of each simulation using the label of $M:P$. $M$ and $P$ illustrate respectively the spatial resolution of the advection-diffusion and electric potential Poisson solvers. When the CGP scheme is applied to both the electric potential and the pressure Poisson equations, $P$ represents the mesh resolution of the pressure Poisson equation as well.

We study the effect of the CGP strategy on the incompressible MHD flow for two different Stuart numbers: $N = 0.01$ and $N = 0.50$. Figure 3 shows the time evolution of the lift $(C_L)$ and drag $(C_D)$ coefficients for these Stuart numbers when the flow field is simulated using the standard algorithm $(k = 0)$ with the spatial resolution of 215680:215680. The magnitude of both the lift and drag forces increase as the Lorentz force increases. Similar observations have been reported by [10]. It is worth noting that this increment is considerable even when $N = 0.01$ in comparison to $N = 0.00$.

We present the numerical results of the current study in two different sections. Section 3.1 gives the outcomes when the CGP method is only applied to the electric potential Poisson solver. Section 3.2; however, provides the results when the CGP algorithm is simultaneously used to execute both the electric potential and pressure Poisson solvers.

### 3.1 Applying CGP to the electric potential Poisson equation

Figure 4 provides a visual comparison between the obtained vorticity fields with and without the CGP method for $N = 0.01$ and $N = 0.50$ at time $t = 150\,\text{s}$. At each row of these figures, we compare the resulting fields with three different resolutions: full fine scale, CGP scale, and full coarse scale. The CGP method provides more detailed data compared to that modeled with a pure coarse grid resolution for all levels of coarsening.

To more precisely investigate the performance of the CGP idea, we plot the vorticity distribution along the horizontal centerline, behind the cylinder and in the wake region at time $t = 150$ s in Fig. 5 for $N = 0.01$ and $N = 0.50$. As can be seen from Fig. 5, the resulting vorticity field for $N = 0.01$ captured from the full fine scale simulation performed on the 215680:215680 grid resolution and the vorticity distribution for all coarsening levels of the CGP simulations are identical. Note that the outputs of the full coarse mesh include spurious fluctuations at the end of domain. Additionally, the fields computed by the CGP and standard high-resolution (215680:215680) algorithms oscillate with the same phase lag. As depicted in Fig. 5 by increasing the Lorentz force ($N = 0.50$), the performance of CGP varies. For one and two levels ($k = 1$ and $k = 2$) of coarsening when $N = 0.50$ (see Fig. 5c and d), although the phases of periodic variation of the CGP and standard high-resolution (215680:215680) schemes are not equal to each other, CGP filters artificial fluctuations contaminating the flow field simulated with the pure coarse resolutions.

An exact measurement of the lift and drag forces are usually needed for engineering designs. In order to examine the efficiency of the CGP framework from this perspective, we plot the time evolution of the lift and drag coefficients of the flow past a cylinder for Stuart numbers of $N = 0.01$ and $N = 0.50$ for different spatial grid resolutions in Fig 6. To save space, we simply exhibit six different situations as a few examples in Fig. 6. In all cases, the predictions of CGP are close to the full fine scale results, while they are significantly more accurate than the full coarse outputs.

The total CPU time consumed by the CGP scheme and the standard algorithm as well as the leading speedup factors for different mesh resolutions and for the Stuart numbers of $N = 0.01$ and $N = 0.50$ are tabulated in Table 1. We restrict the data to the case of well-resolved solutions. Hence, the data corresponded to the three levels ($k = 3$) of grid coarsening is not shown in Table 1. Accordingly, the maximum achieving speed up factor is 3.549. In general, the accelerating factors obtained for $N = 0.01$ are higher than for $N = 0.50$ in the equivalent CGP resolution, as by increasing the Lorentz force the resulting system of equations becomes stiffer.

Figure 7 compares the CPU times consumed by various components of the problem for the Stuart number of $N = 0.01$. As can be seen from Fig. 7, the electric potential Poisson equation is the most time consuming component of the process when the standard approach ($k=0$) is used. Surprisingly, this cost overcomes even the computational price devoted to the pressure Poisson equation. By the coarsening level increment, the electric potential Poisson equation expense lessens dramatically so that its portion becomes less than 1% after just two levels of coarsening. For three levels of coarsening, the electric potential Poisson solver cost becomes insignificant, while the pressure Poisson equation consumes the majority of the computational resources. That is our motivation to apply the CGP technique to the pressure Poisson equation in order to obtain the maximum possible speedup factor for the simulation of incompressible MHD flows.

### 3.2 Applying CGP to the electric potential Poisson equation and the pressure Poisson equation

As we discussed in the previous section, the CGP configuration is proficient in order to reduce the computational cost associated with the electric potential Poisson equation while it preserves the accuracy of the velocity field in the presence of an external magnetic field. On the other hand, a considerable number of researchers in the literature [17, 23, 14, 11, 13, 12] demonstrated the efficiency of the CGP algorithm when it was applied to the solution of the pressure Poisson equation in pressure correction schemes to numerically solve the incompressible Navier-Stokes equations. The main concern of this section is to assess the capability of CGP in terms of both accelerating the computations and preserving the accuracy of the velocity field when one applies CGP to both the pressure Poisson and electric potential Poisson equations simultaneously in the numerical simulations of incompressible MHD fields. Similar to the previous section, we consider simulations with Stuart numbers of $N=0.01$ and $N=0.50$. But to save space, we mainly present the results of $N=0.01$.

Figure 8 exposes the vorticity field computed by the standard algorithm ($k=0$) and the CGP scheme for $k=1, 2,$ and $3$ for the Stuart number of $N=0.01$ at time $t=150\text{s}$. From a visual point of view, the fidelity of the CGP vorticity fields are preserved for one and two levels of the grid coarsening ($k=1$, and $2$) and they are close to the full fine scale mesh resolution ($k=0$, 215680:215680). More notably, the CGP vorticity fields for $k=1$ (215680: 53920) and $k=2$ (215680: 13480) have a higher level of accuracy compared to the results obtained by the corresponding full coarse resolutions (e.g., 53920:53920 and 13480:13480). For three coarsening levels ($k=3$); however, the fidelity of the vorticity field computed by a CGP simulation with the

215680:3370 resolution is only preserved in a reasonable range but yet significantly provides more accurate information compared to the simulations performed on the coarse scale resolution of 3370:3370.

As can be seen from Fig. 9, we realize that at some cases the CGP methodology leads to a phase lag between the outputs with 215680:215680 and 215680:53920 or 215680:13480 grid resolutions. Our numerical experiments express that these types of phase lags between the outputs of the CGP and standard mechanisms also depend on the time increment ($\delta t$) chosen. Similar observations are reported by Kashefi and Staples [14] when they studied the CGP effect on the simulation of the incompressible flow past a cylinder. Notwithstanding this effect, the key fact is that although dampened flows around the cylinder surfaces and spurious fluctuations at the end of the domain can be observed in the numerical results of the corresponding coarse scale, they are completely removed by the CGP strategy as shown in Fig. 9.

Lift and drag are both two critical engineering quantities in the industry. Thus is it vital to validate the CGP performance for the computations of these two forces. In Sect. 3.1 of this article, we examined the CGP platform for the total lift and drag calculations when we applied CGP only to the electric potential Poisson equation. Kashefi and Staples [14] applied CGP to the pressure Poisson equation to simulate incompressible flows past a circular cylinder. Through their numerical outcomes, they showed that the pressure drag/lift forces have a lower sensitivity to the grid resolution in comparison with the viscous drag/lift forces (e.g., see Fig. 11 in Ref. [14]). The same experience is repeated here, when we applied the CGP methodology to both the pressure Poisson and electric potential Poisson equations. With this in mind, we display the time evolution of the viscous drag and viscous lift for the Stuart number of $N = 0.01$ in Fig. 10. With reference to Fig. 11, the viscous drag and lift coefficients computed by the CGP algorithm are more accurate than those coefficients obtained by standard coarse scale simulations. Specifically, the CGP performance for the viscous drag coefficient ($C_{Df}$) calculation is significantly higher than the full coarse scale computation as can be seen from Fig. 10. For instance, we observe from Fig. 10(d) that although the output of CGP with three coarsening level ($k = 3$, 215680:3370) has 2.3% error in the viscous drag coefficient with reference to the standard full fine scale computation (215680:215680), the error of full coarse scale simulation (3370:3370) is 34.5% in respect to the same reference.

So far, we investigated the accuracy of the velocity and vorticity fields and the viscous drag and lift forces computed by CGP. Another important aspect of the CGP applications is speeding up the computations. Table 2 and Figure 11 allocate the relevant information.

The information collected in Table 2 demonstrates the CPU time of CGP and non-CGP simulations along with their achieved speedups. Similar to Table 1, we only present the data relevant to the well-resolved solutions in Table 2. In contrast with the trend tabulated in Table 1, by increasing the Lorentz force the speedup factor increases. In this case, the maximum speedup is approximately a factor of 23. As expected we experience higher levels of accelerations in comparison with the data tabulated in Table 1, as CGP is right now applied to both the electric potential Poisson and the pressure Poisson equations.

As can be observed in Fig. 11, the portion of computational expenses associated with both the pressure Poisson and the electric potential Poisson equations becomes less than 7% only by means of two levels of coarsening. According to the information provided in Fig. 11, the electric potential Poisson equation is stiffer than the pressure Poisson equation even after three levels of coarsening ($k = 3$).

## 4 Conclusions

In the present work, the CGP multiresolution algorithm was used for the first time to obtain a numerical solution to incompressible MHD flows. CGP saved the computational resources mainly due to solving the electric potential Poisson equation on a coarser grid relative to that one used for the simulation of the advection-diffusion equation. The maximum speed up in this case was approximately a factor of 3. To gain a higher level of speed up, we utilized CGP for the pressure Poisson equation in addition. This strategy led to 23 times speedup, maximum. The CGP performance was investigated by the examination of the structure of the von Karman street for the vorticity field, the velocity field in the wake region, and the drag and lift coefficient evolution. Our numerical experiments demonstrated that CGP was able to maintain the accuracy of quantities of interest with reference to standard algorithms with high resolution grids for both the advection-diffusion and the electric potential Poisson equation, while reduced computational expenses dramatically. Notably, the CGP results were significantly more accurate than standard algorithms with low resolution grids for both the above mentioned equations.

**Acknowledgments**

AK wishes to thank Dr. Peter Minev for helpful discussions.

**References**

[1]  A. H. BOOZER, *Ohm's law for mean magnetic fields*, Journal of plasma physics, 35 (1986), pp. 133-139.
[2]  A. J. CHORIN, *Numerical solution of the Navier-Stokes equations*, Mathematics of computation, 22 (1968), pp. 745-762.
[3]  V. DOUSSET and A. POTHÉRAT, *Numerical simulations of a cylinder wake under a strong axial magnetic field*, Physics of Fluids, 20 (2008), pp. 017104.
[4]  U. FAROOQ, D. LU, S. MUNIR, M. RAMZAN, M. SULEMAN and S. HUSSAIN, *MHD flow of Maxwell fluid with nanomaterials due to an exponentially stretching surface*, Scientific reports, 9 (2019), pp. 7312.
[5]  N. GANTA, B. MAHATO and Y. G. BHUMKAR, *Analysis of sound generation by flow past a circular cylinder performing rotary oscillations using direct simulation approach*, Physics of Fluids, 31 (2019), pp. 026104.
[6]  C. GEUZAINE and J. F. REMACLE, *Gmsh: A 3-D finite element mesh generator with built-in pre-and post-processing facilities*, International Journal for Numerical Methods in Engineering, 79 (2009), pp. 1309-1331.
[7]  S. GHOSH, S. SARKAR, R. SIVAKUMAR and T. SEKHAR, *Forced convection magnetohydrodynamic flow past a circular cylinder by considering the penetration of magnetic field inside it*, Numerical Heat Transfer, Part A: Applications (2019), pp. 1-18.
[8]  M. IRFAN and M. A. FAROOQ, *Magnetohydrodynamic free stream and heat transfer of nanofluid flow over an exponentially radiating stretching sheet with variable fluid properties*, Frontiers in Physics, 7 (2019), pp. 186.
[9]  M. JIN, W. LIU and Q. CHEN, *Accelerating fast fluid dynamics with a coarse-grid projection scheme*, HVAC&R Research, 20 (2014), pp. 932-943.
[10] J. JOSSERAND, P. MARTY and A. ALEMANY, *Pressure and drag measurements on a cylinder in a liquid metal flow with an aligned magnetic field*, Fluid dynamics research, 11 (1993), pp. 107.
[11] A. KASHEFI, *A coarse-grid incremental pressure projection method for accelerating low Reynolds number incompressible flow simulations*, Iran Journal of Computer Science (2018), pp. 1-11.
[12] A. KASHEFI, *A Coarse Grid Projection Method for Accelerating Free and Forced Convection Heat Transfer Computations*, Results in Mathematics, 75 (2020), pp. 33.
[13] A. KASHEFI, *Coarse grid projection methodology: A partial mesh refinement tool for incompressible flow simulations*, Bulletin of the Iranian Mathematical Society (2018), pp. 1-5.
[14] A. KASHEFI and A. E. STAPLES, *A finite-element coarse-grid projection method for incompressible flow simulations*, Advances in Computational Mathematics, 44 (2018), pp. 1063-1090.
[15] R. KLEIN and A. POTHÉRAT, *Appearance of three dimensionality in wall-bounded MHD flows*, Physical review letters, 104 (2010), pp. 034502.

[16] H. LEE, M. HA and H. YOON, *A numerical study on the fluid flow and heat transfer in the confined jet flow in the presence of magnetic field*, International journal of heat and mass transfer, 48 (2005), pp. 5297-5309.

[17] M. LENTINE, W. ZHENG and R. FEDKIW, *A novel algorithm for incompressible flow using only a coarse grid projection*, ACM Transactions on Graphics (TOG), ACM, 2010, pp. 114.

[18] B. MÜCK, C. GÜNTHER, U. MÜLLER and L. BÜHLER, *Three-dimensional MHD flows in rectangular ducts with internal obstacles*, Journal of Fluid Mechanics, 418 (2000), pp. 265-295.

[19] G. MUTSCHKE, G. GERBETH, V. SHATROV and A. TOMBOULIDES, *Two-and three-dimensional instabilities of the cylinder wake in an aligned magnetic field*, Physics of Fluids, 9 (1997), pp. 3114-3116.

[20] Y. PENG, A. S. ALSAGRI, M. AFRAND and R. MORADI, *A numerical simulation for magnetohydrodynamic nanofluid flow and heat transfer in rotating horizontal annulus with thermal radiation*, RSC advances, 9 (2019), pp. 22185-22197.

[21] J. N. REDDY, *An introduction to the finite element method*, McGraw-Hill New York, 1993.

[22] Y. SAAD and M. H. SCHULTZ, *GMRES: A generalized minimal residual algorithm for solving nonsymmetric linear systems*, SIAM Journal on scientific and statistical computing, 7 (1986), pp. 856-869.

[23] O. SAN and A. E. STAPLES, *A coarse-grid projection method for accelerating incompressible flow computations*, Journal of Computational Physics, 233 (2013), pp. 480-508.

[24] O. SAN and A. E. STAPLES, *An efficient coarse grid projection method for quasigeostrophic models of large-scale ocean circulation*, International Journal for Multiscale Computational Engineering, 11 (2013).

[25] Z. SHAH, H. BABAZADEH, P. KUMAM, A. SHAFEE and P. THOUNTHONG, *Numerical simulation of magnetohydrodynamic nanofluids under the influence of shape factor and thermal transport in a porous media using CVFEM*, Front. Phys. 7: 164. doi: 10.3389/fphy (2019).

[26] M. SHOAIB, R. AKHTAR, M. A. R. KHAN, M. A. RANA, A. M. SIDDIQUI, Z. ZHIYU and M. A. Z. RAJA, *A Novel Design of Three-Dimensional MHD Flow of Second-Grade Fluid past a Porous Plate*, Mathematical Problems in Engineering, 2019 (2019).

[27] J. VERRON and J. SOMMERIA, *Numerical simulation of a two-dimensional turbulence experiment in magnetohydrodynamics*, The Physics of fluids, 30 (1987), pp. 732-739.

[28] H. YOON, H. CHUN, M. HA and H. LEE, *A numerical study on the fluid flow and heat transfer around a circular cylinder in an aligned magnetic field*, International Journal of Heat and Mass Transfer, 47 (2004), pp. 4075-4087.

[29] M. ZHAO and L. LU, *Numerical simulation of flow past two circular cylinders in cruciform arrangement*, Journal of Fluid Mechanics, 848 (2018), pp. 1013-1039.

**Table 1** Comparison of total CPU times between the standard and CGP algorithms for two different Stuart numbers when the CGP solver is only used for the electric potential Poisson equation.

| k | Resolution | $N = 0.01$ | | $N = 0.50$ | |
|---|---|---|---|---|---|
| | | CPU (s) | Speedup | CPU (s) | Speedup |
| 0 | 215680:215680 | 3148020.0 | 1.000 | 3711894.0 | 1.000 |
| 1 | 215680:53920 | 957878.0 | 3.286 | 1272233.0 | 2.917 |
| 2 | 215680:13480 | 886935.0 | 3.549 | 1214636.0 | 3.055 |
| 0 | 53920:53920 | 252062.0 | 1.000 | 150768.0 | 1.000 |
| 0 | 13480:13480 | 25561.7 | 1.000 | 16867.6 | 1.000 |

**Table 2** Comparison of total CPU times between the standard and CGP algorithms for two different Stuart numbers when the CGP solver is applied to both the electric potential Poisson equation and the pressure Poisson equation

| k | Resolution | $N = 0.01$ CPU (s) | Speedup | $N = 0.50$ CPU (s) | Speedup |
|---|---|---|---|---|---|
| 0 | 215680:215680 | 3148020.0 | 1.000 | 3711894.0 | 1.000 |
| 1 | 215680:53920 | 280419.0 | 11.226 | 275930.0 | 13.452 |
| 2 | 215680:13480 | 154402.0 | 20.388 | 159979.0 | 23.202 |
| 0 | 53920:53920 | 252062.0 | 1.000 | 150768.0 | 1.000 |
| 0 | 13480:13480 | 25561.7 | 1.000 | 16867.6 | 1.000 |

**Fig. 1** Scheme of coarse grid projection methodology involving the restriction and prolongation of the velocity and electric potential data. Some parts of this figure are reproduced from Fig. 1 of Ref. [14].

**Fig. 2** Computational grid for the electric potential Poisson equation and the pressure Poisson equations **a** after two levels coarsening $(k=2)$; **b** after three levels coarsening $(k=3)$. This figure is reproduced from Ref. [14].

**Fig. 3** A Comparison between the time evolution of **a** lift coefficient for Stuart numbers of $N=0.00$, $N=0.01$, and $N=0.50$, and **b** drag coefficient for Stuart numbers of $N=0.00$, $N=0.01$, and $N=0.50$. For all the test cases, the standard algorithm $(k=0)$ is used.

**Fig. 4** Vorticity fields at $t=150\text{s}$ for the Stuart number of $N=0.01$ and $N=0.50$ when the CGP solver is only used for the electric potential Poisson equation.

**Fig. 5** Comparison of vorticity in the wake of the flow over a cylinder at $t=150\text{s}$ when the CGP solver is only used for the electric potential Poisson equation for different values of the electric potential Poisson, the advection-diffusion, and the pressure Poisson grid resolutions. **a** Vorticity for the Stuart number of $N=0.01$ for CGP $(k=0, 2, \text{and } 0)$; **b** Vorticity for the Stuart number of $N=0.01$ for CGP $(k=0, 3, \text{and } 0)$; **c** Vorticity for the Stuart number of $N=0.50$ for CGP $(k=0, 1, \text{and } 0)$; **d** Vorticity for the Stuart number of $N=0.50$ for CGP $(k=0, 2, \text{and } 0)$.

**Fig. 6** A Comparison between the time evolution of lift and drag coefficients when the CGP solver is only used for the electric potential Poisson equation for different values of the electric potential Poisson, the advection-diffusion, and the pressure Poisson grid resolutions. **a** Drag coefficient for the Stuart number of $N=0.01$ for CGP $(k=0, 1, \text{and } 0)$; **b** Drag coefficient for the Stuart number of $N=0.01$ for CGP $(k=0, 2, \text{and } 0)$; **c** Drag coefficient for the Stuart number of $N=0.01$ for CGP $(k=0, 3, \text{and } 0)$; **d** Lift coefficient for the Stuart number of $N=0.01$ for CGP $(k=0, 3, \text{and } 0)$; **e** Drag coefficient for the Stuart number of $N=0.50$ for

CGP $(k = 0, 1, \text{ and } 0)$; **f** Drag coefficient for the Stuart number of $N = 0.50$ for CGP $(k = 0, 2, \text{ and } 0)$.

**Fig. 7** Percent CPU times for the electric potential Poisson part, the advection–diffusion part, and the pressure Poisson part for the standard and CGP algorithms for the Stuart number of $N = 0.01$ when the CGP solver is only used for the electric potential Poisson equation. For all the test cases, the mapping cost is less than 0.001% and is not shown.

**Fig. 8** Vorticity fields at $t = 150 \text{s}$ for the Stuart number of $N = 0.01$ when the CGP solver is applied to both the electric potential Poisson equation and the pressure Poisson equation.

**Fig. 9** Comparison of vorticity in the wake of the flow over a cylinder at $t = 150 \text{s}$ when the CGP solver is applied to both the electric potential Poisson equation and the pressure Poisson equation for different values of the electric potential Poisson, the advection-diffusion, and the pressure Poisson grid resolutions. **a** Vorticity for the Stuart number of $N = 0.01$ for CGP $(k = 0, 1, \text{ and } 0)$; **b** Vorticity for the Stuart number of $N = 0.01$ for CGP $(k = 0, 2, \text{ and } 0)$.

**Fig. 10** A Comparison between the time evolution of viscous lift and viscous drag coefficients for the Stuart number of $N = 0.01$ when the CGP solver is applied to both the electric potential Poisson equation and the pressure Poisson equation for different values of the electric potential Poisson, the advection-diffusion, and the pressure Poisson grid resolutions. **a** Viscous lift coefficient for CGP $(k = 0, 1, \text{ and } 0)$; **b** Viscous drag coefficient for CGP $(k = 0, 1, \text{ and } 0)$; **c** Viscous lift coefficient for CGP $(k = 0, 2, \text{ and } 0)$; **d** Viscous drag coefficient for CGP $(k = 0, 2, \text{ and } 0)$; **e** Viscous lift coefficient for CGP $(k = 0, 3, \text{ and } 0)$; **f** Viscous drag coefficient for CGP $(k = 0, 3, \text{ and } 0)$.

**Fig. 11** Percent CPU times for the electric potential Poisson part, the advection–diffusion part, and the pressure Poisson part for the standard and CGP algorithms for the Stuart number of $N = 0.01$ when the CGP solver is applied to both the electric potential Poisson equation and the pressure Poisson equation. For all the test cases, the mapping cost is less than 0.001% and is not shown.

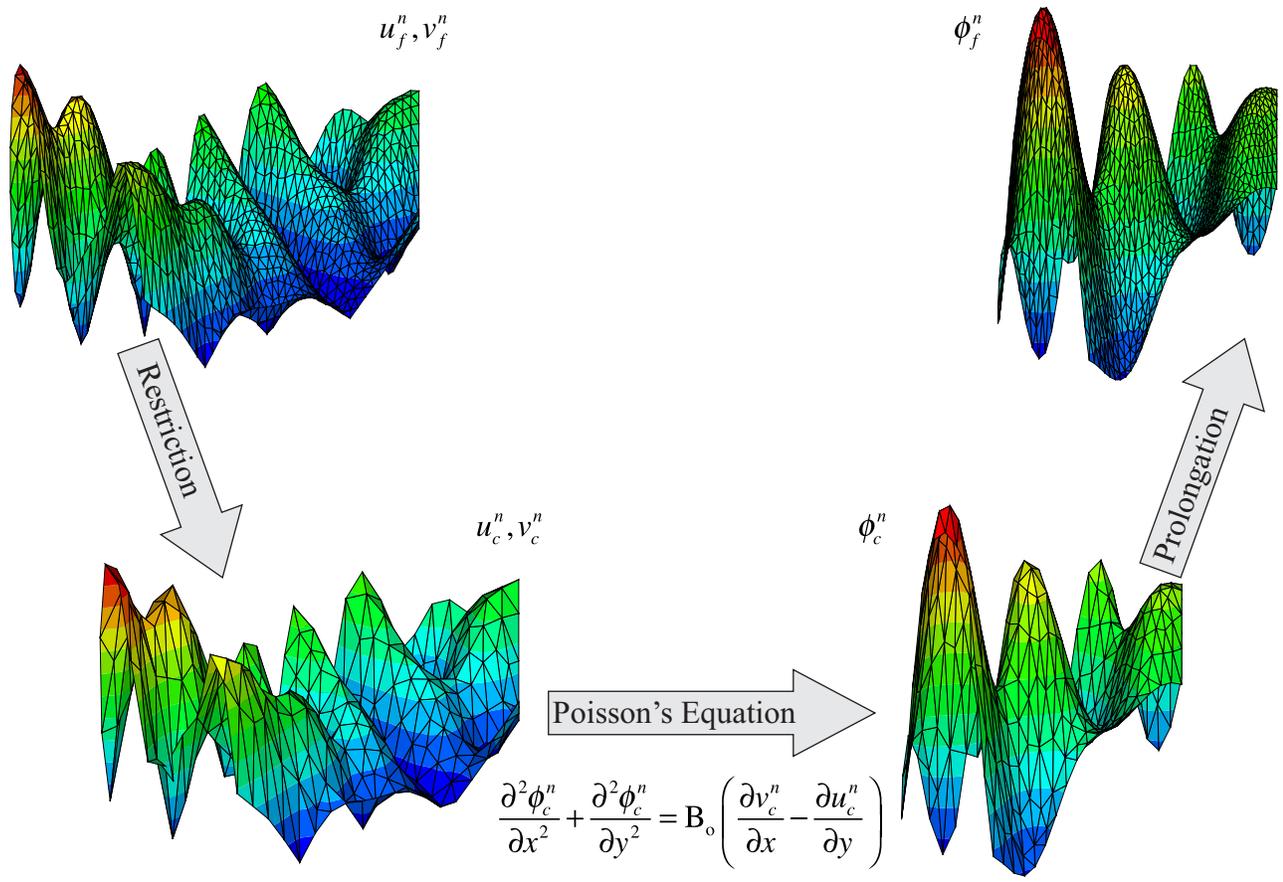

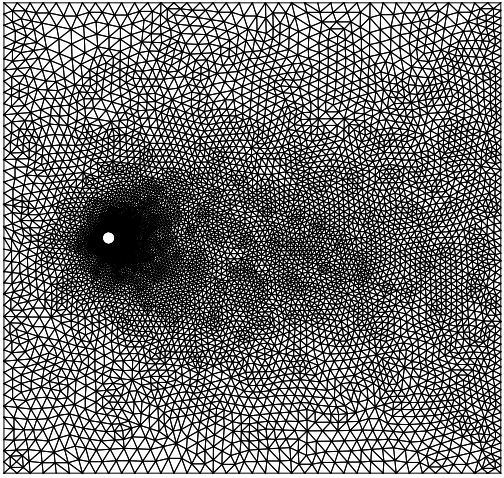 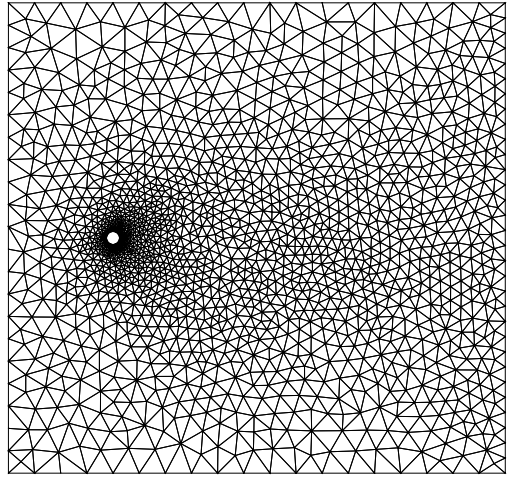

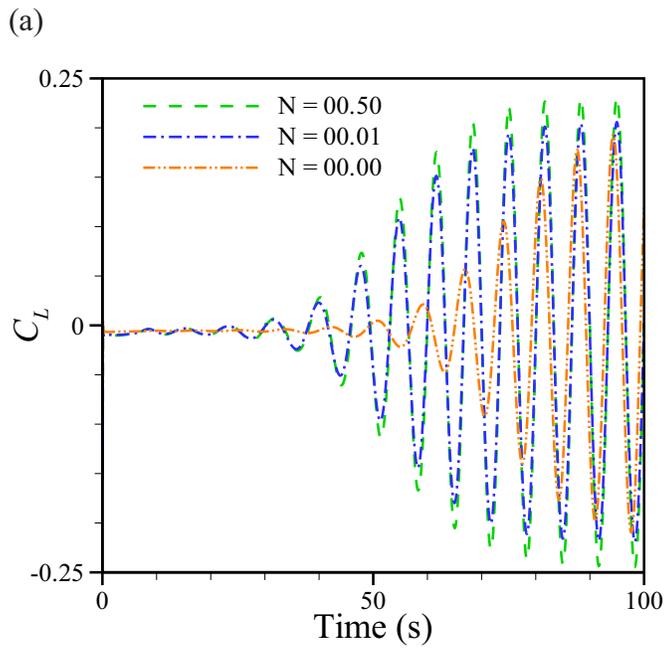 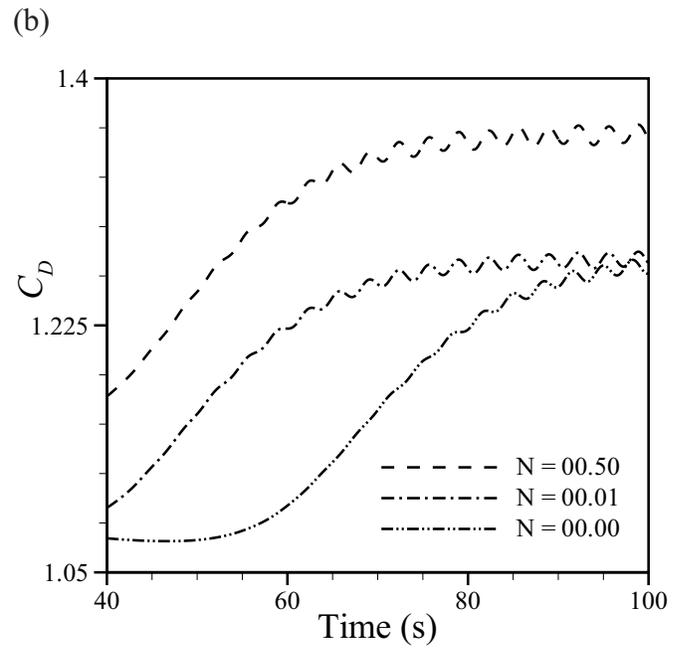

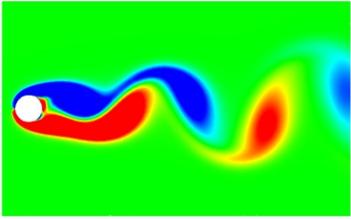 215680:215680 N = 0.01
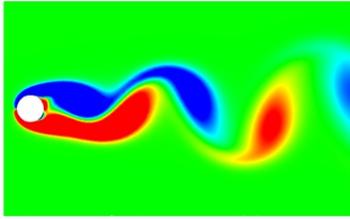 215680:53920 N = 0.01
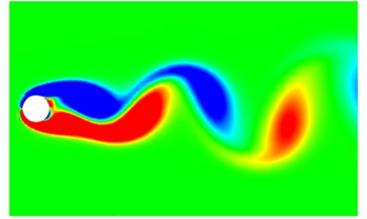 53920:53920 N = 0.01

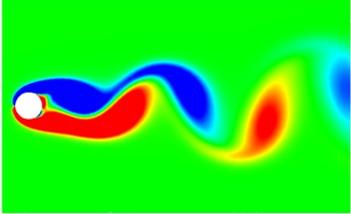 215680:215680 N = 0.01
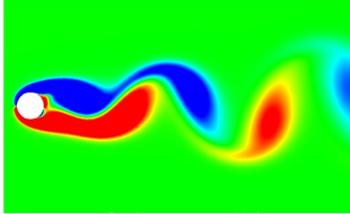 215680:13480 N = 0.01
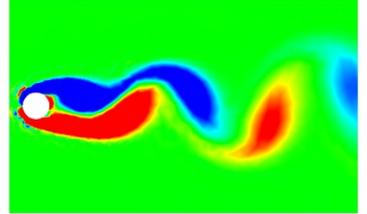 13480:13480 N = 0.01

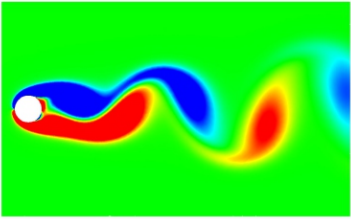 215680:215680 N = 0.01
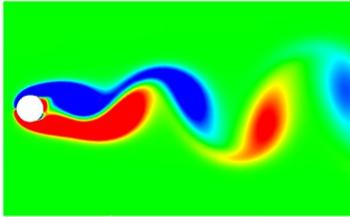 215680:3370 N = 0.01
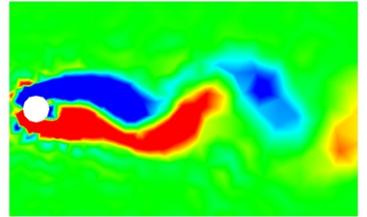 3370:3370 N = 0.01

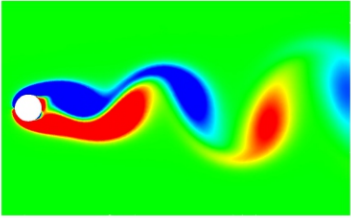 215680:215680 N = 0.50
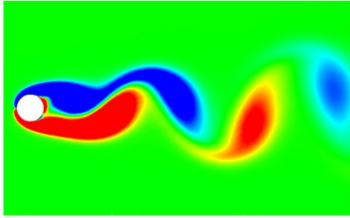 215680:53920 N = 0.50
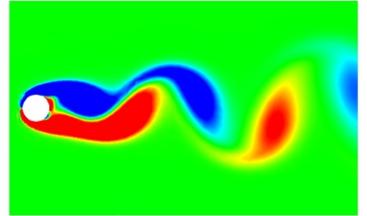 53920:53920 N = 0.50

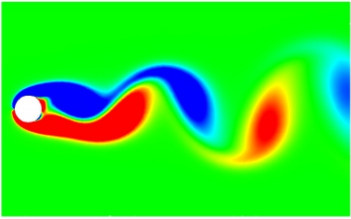 215680:215680 N = 0.50
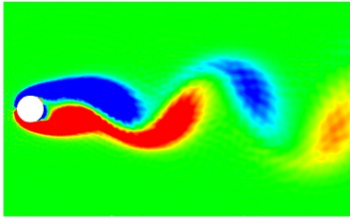 215680:13480 N = 0.50
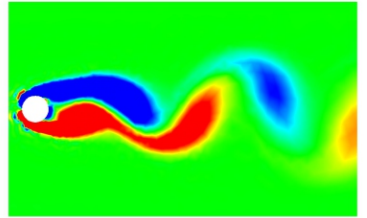 13480:13480 N = 0.50

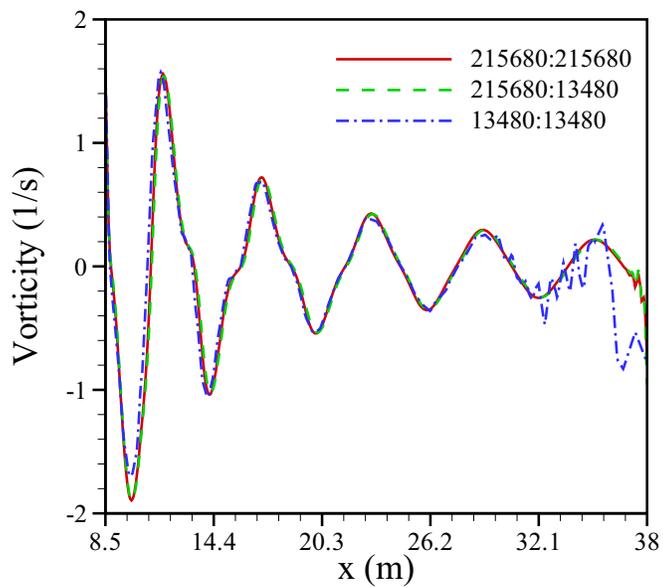
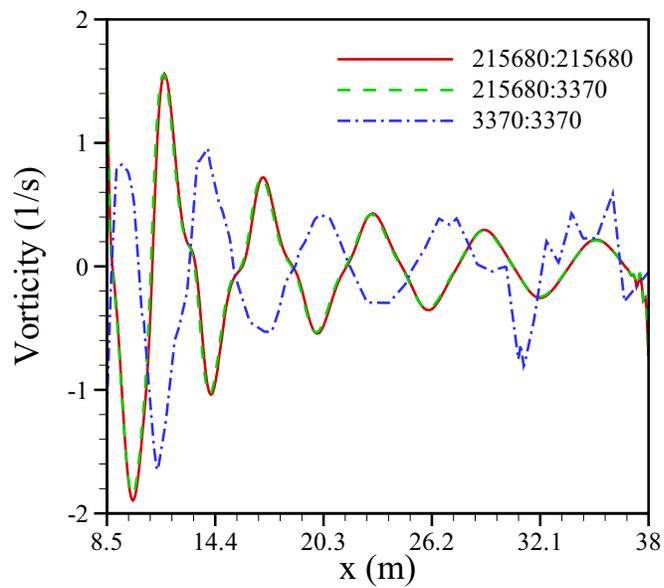
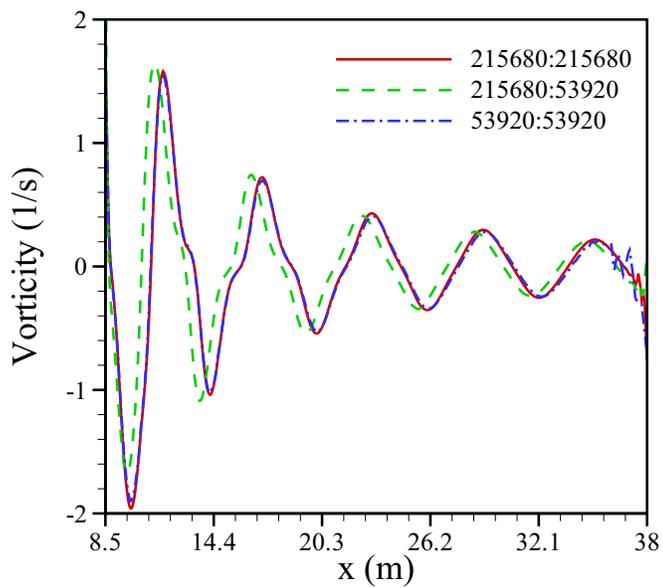
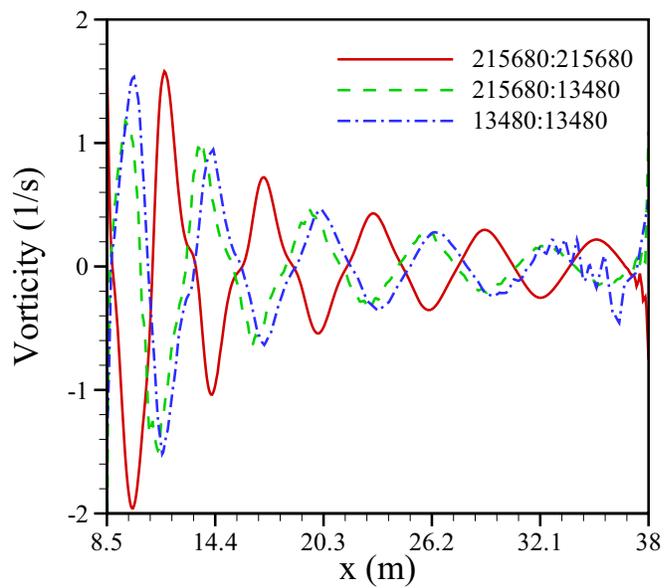

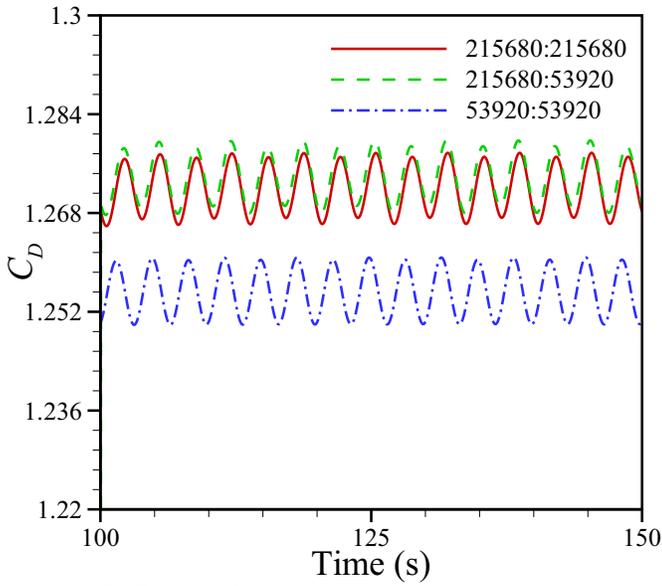
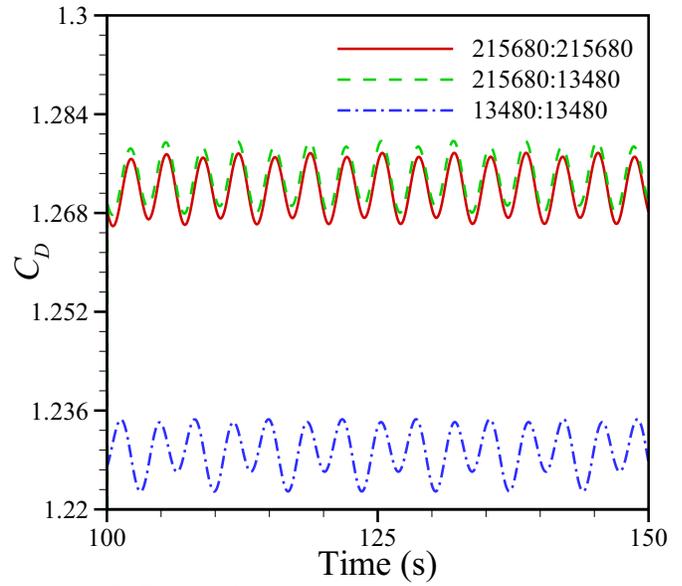
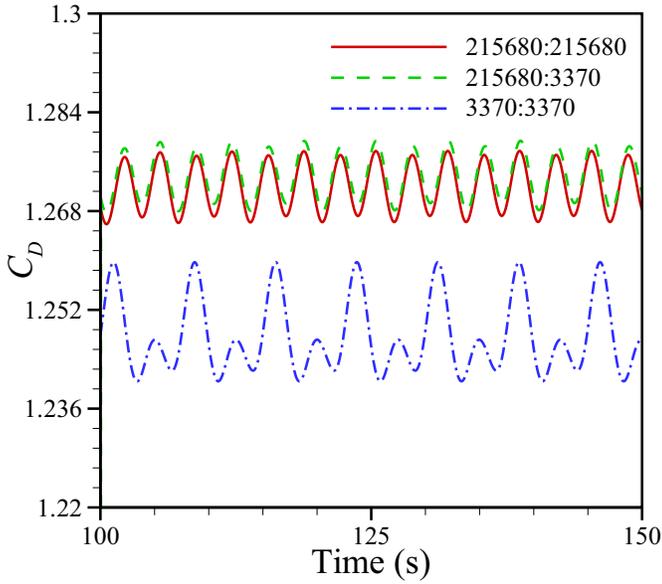
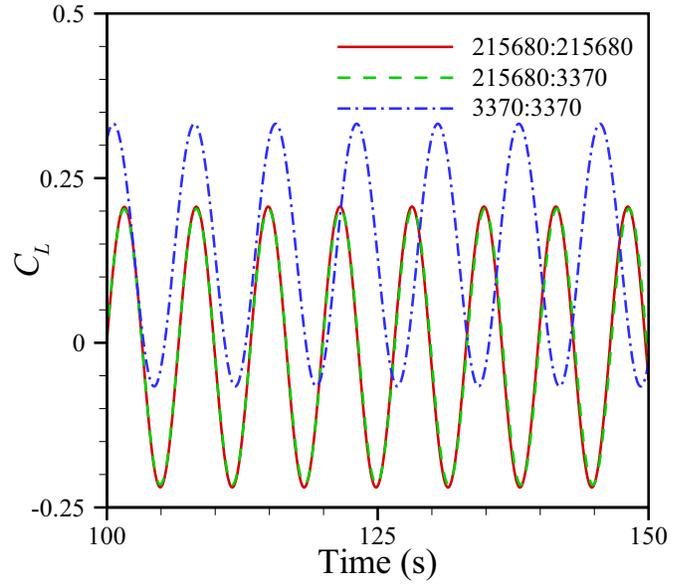
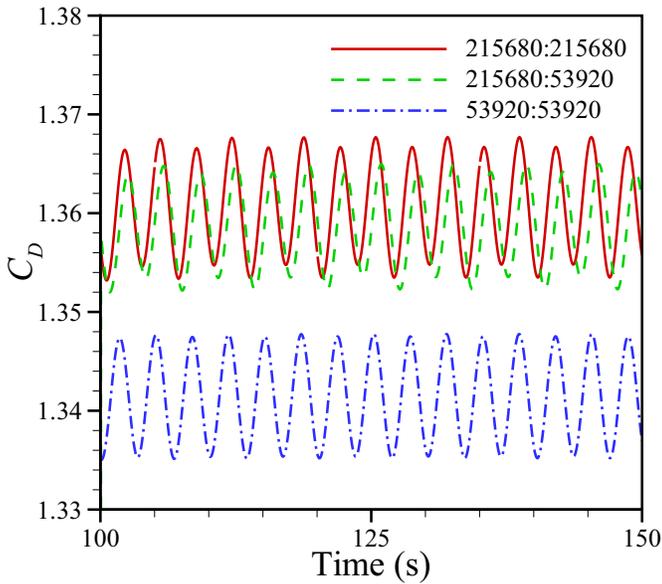
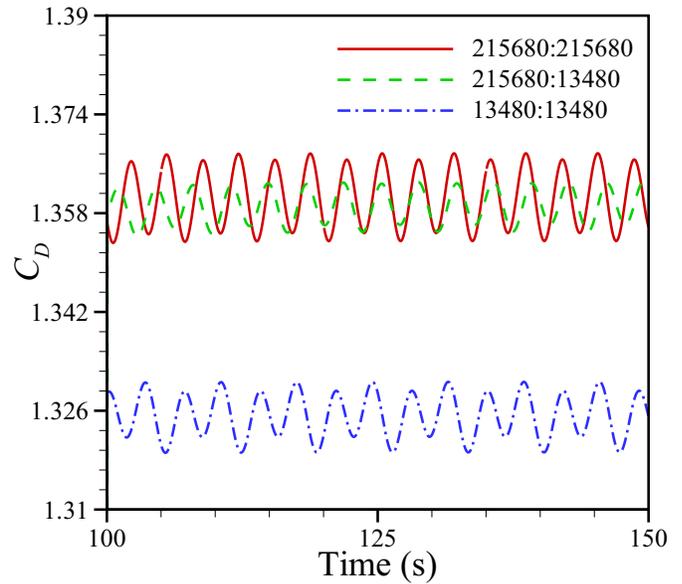

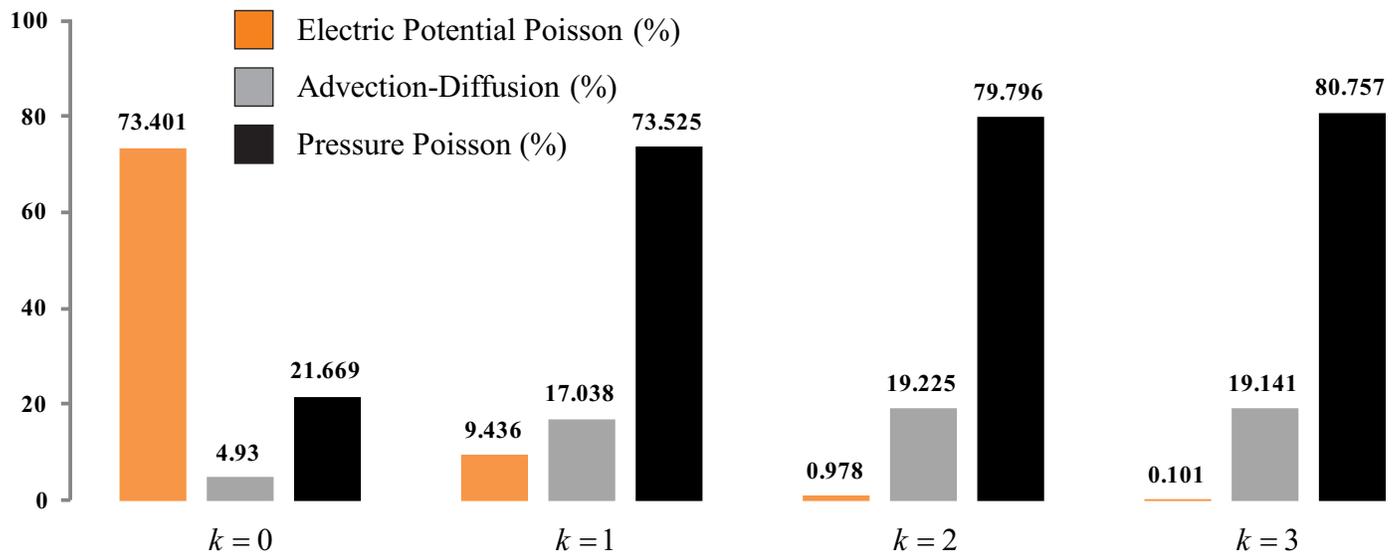

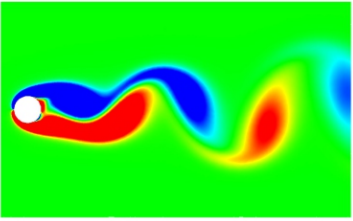 215680:215680
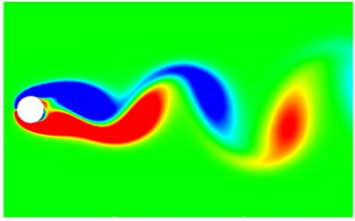 215680:53920
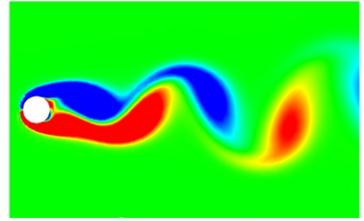 53920:53920

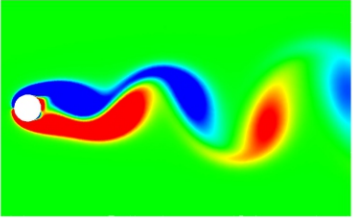 215680:215680
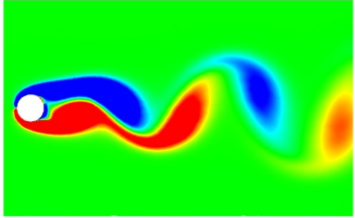 215680:13480
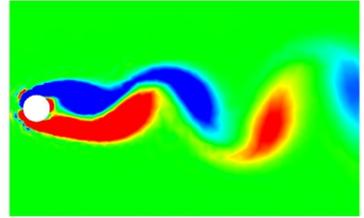 13480:13480

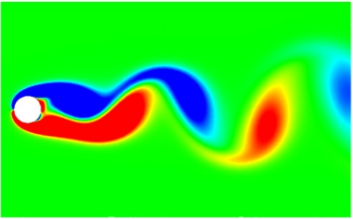 215680:215680
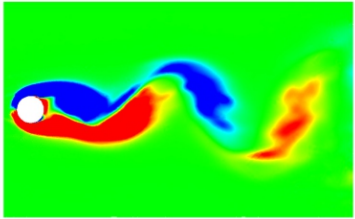 215680:3370
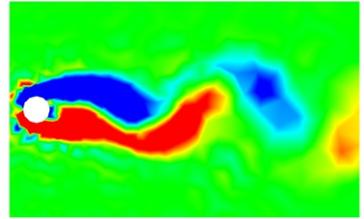 3370:3370

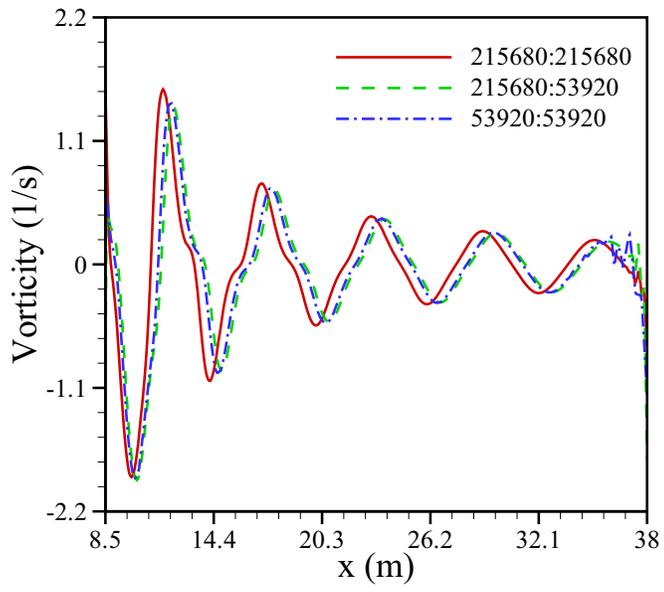 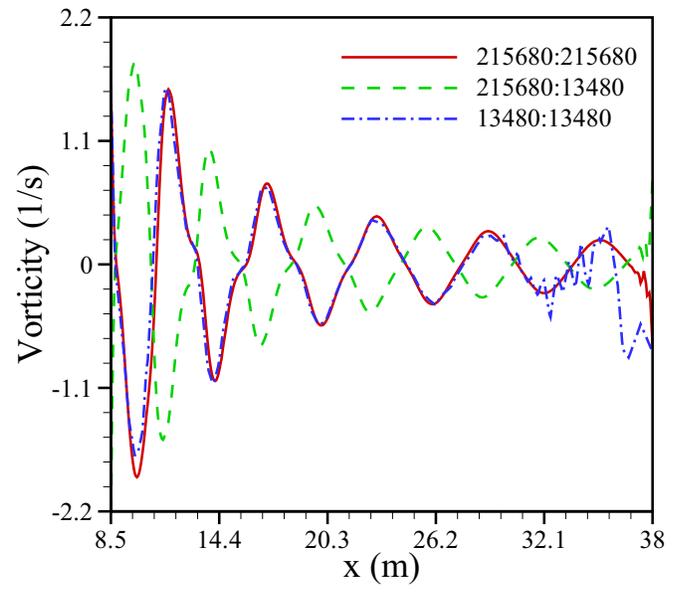

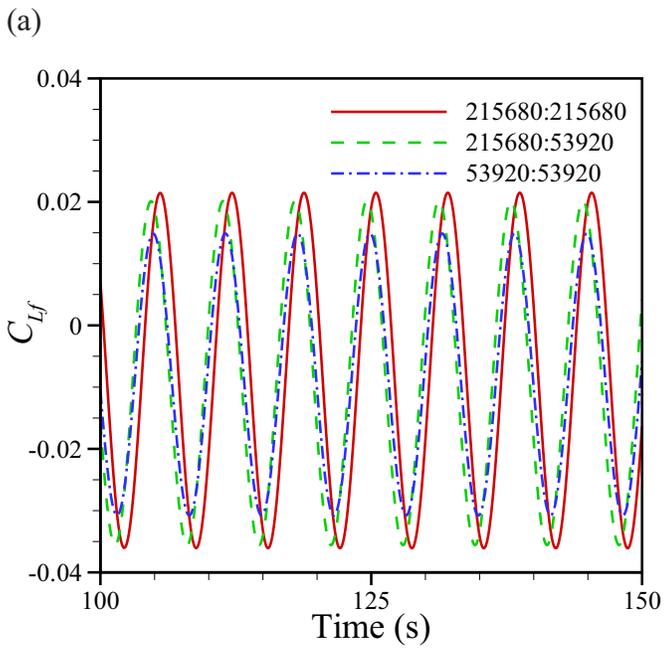
(a)

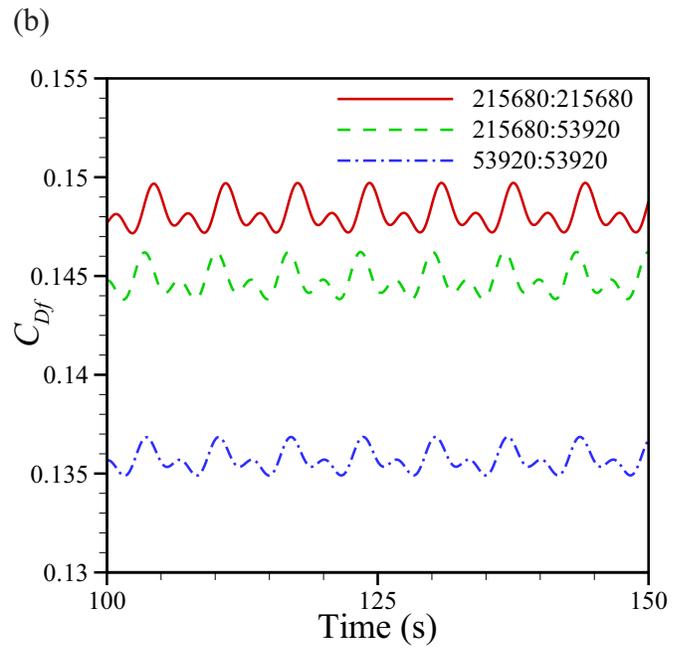
(b)

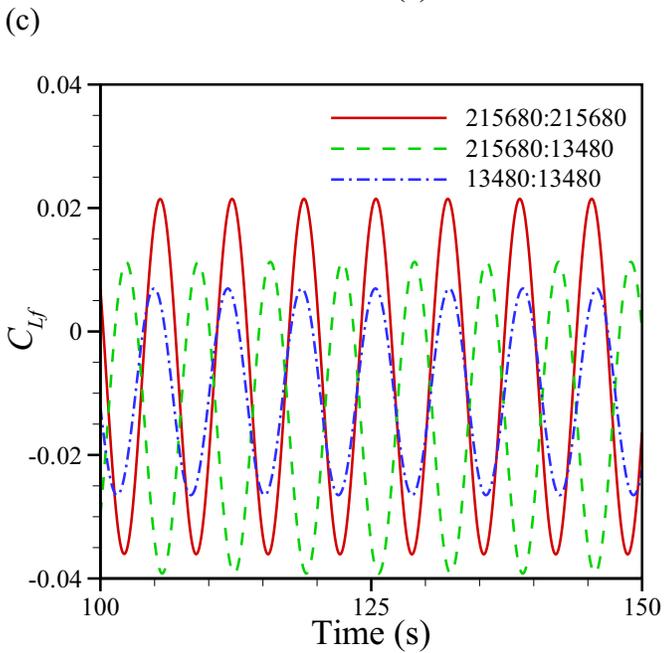
(c)

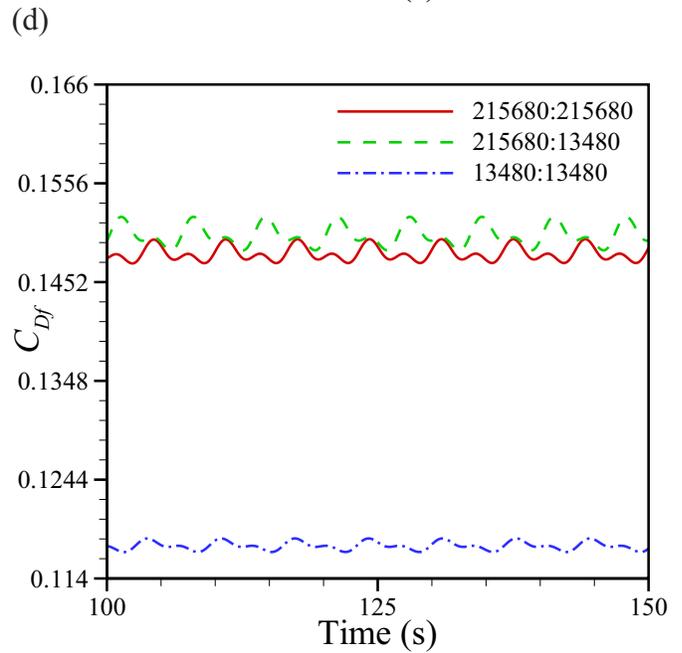
(d)

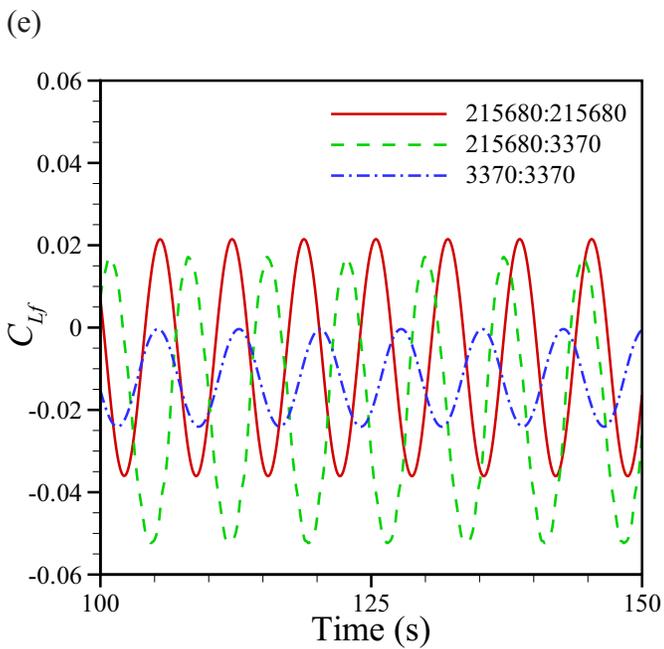
(e)

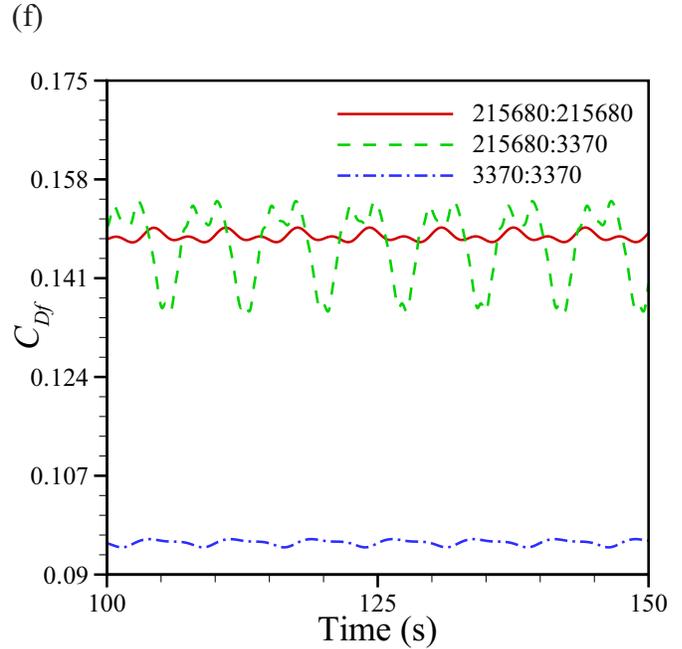
(f)

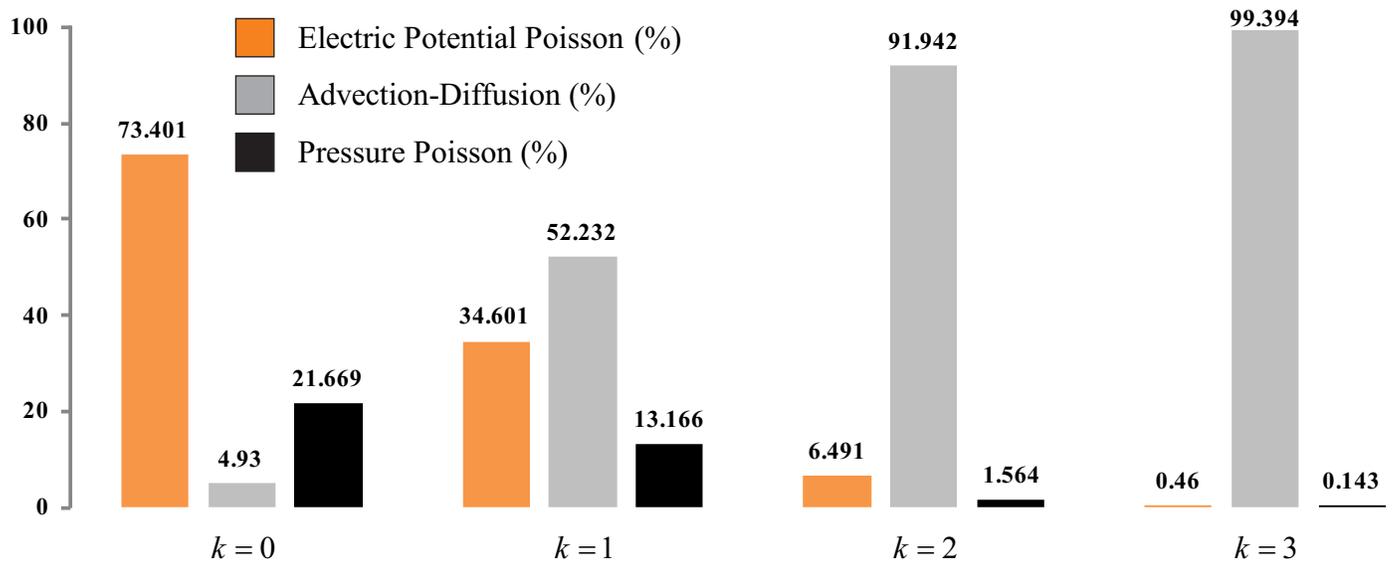